\documentclass[aps, prl, preprint, amsmath, amssymb, floatfix, superscriptaddress]{revtex4-2}

\usepackage{graphicx} 
\usepackage{dcolumn}  
\usepackage{bm}       
\usepackage{float}
\usepackage{gensymb}
\usepackage{xcolor}
\usepackage{units}
\usepackage{placeins}
\usepackage{ulem}
\usepackage{hyperref}
\usepackage{bbm}
\usepackage{multirow}
\usepackage[T1]{fontenc}
\usepackage{xspace}
\usepackage[colorinlistoftodos, color=green!40, prependcaption]{todonotes}
\usepackage{caption}
\def\Neel{N\'{e}el\xspace}
\newcommand{\Lvec}{\ensuremath{\mathbf{L}}\xspace}

\begin{document}

\title{Coherent Magnons Driven by Photomodulated Anisotropy in Altermagnetic MnTe}

\author{Dingbin Huang}
\thanks{These two authors contributed equally to this work.}
\affiliation{Materials Science Division, Lawrence Berkeley National Laboratory, Berkeley, CA, USA.}

\author{Jonathon Kruppe}
\thanks{These two authors contributed equally to this work.}
\affiliation{Department of Physics, University of California, Berkeley, CA, USA.}
\affiliation{Materials Science Division, Lawrence Berkeley National Laboratory, Berkeley, CA, USA.}

\author{Resham Babu Regmi}
\affiliation{Department of Physics \& Astronomy, University of Notre Dame, Notre Dame, Indiana 46556, USA}
\affiliation{Stavropoulos Center for Complex Quantum Matter,
University of Notre Dame, Notre Dame, IN, USA 46556}

\author{Nirmal J. Ghimire}
\affiliation{Department of Physics \& Astronomy, University of Notre Dame, Notre Dame, Indiana 46556, USA}
\affiliation{Stavropoulos Center for Complex Quantum Matter,
University of Notre Dame, Notre Dame, IN, USA 46556}

\author{James Analytis}
\affiliation{Department of Physics, University of California, Berkeley, CA, USA.}
\affiliation{Materials Science Division, Lawrence Berkeley National Laboratory, Berkeley, CA, USA.}

\author{Joseph Orenstein}
\email{jworenstein@lbl.gov}
\affiliation{Department of Physics, University of California, Berkeley, CA, USA.}
\affiliation{Materials Science Division, Lawrence Berkeley National Laboratory, Berkeley, CA, USA.}

\begin{abstract}
Manganese telluride ($\text{MnTe}$) has recently emerged as a prototypical $g$-wave altermagnet, providing an ideal platform to investigate the non-equilibrium excitations of altermagnetic order. Here, we report simultaneous spatial mapping of the local equilibrium orientation of the N{\'e}el vector, $\varphi_L(\mathbf{r})$, alongside the amplitude, $\Delta\varphi(\mathbf{r},t)$, and frequency, $\Omega(\mathbf{r})$, of photoexcited spin waves. Based on these measurements, we place a remarkably low 
upper bound of $\approx 60~\mu\text{eV}$ (0.7~K) on the spin-wave gap arising from intrinsic anisotropy. This exceptionally weak hexagonal anisotropy ($K_6$) renders the altermagnetic order highly susceptible to optical tuning, allowing coherent 
spin waves to be driven by a photoinduced enhancement of $K_6$. Above a threshold pump fluence, our spatial maps reveal that this photomodulation manifests as a 
six-fold symmetric sawtooth dependence of $\Delta\varphi$ on $\varphi_L$ and a cycloid-like modulation of $\Omega$. Ultimately, the near-isotropy of the N{\'e}el vector in $\text{MnTe}$ enables optical and mechanical control over the 
orientation of spin-splitting in the electronic band structure, offering new pathways for altermagnetic spintronics.
\end{abstract}
\maketitle


\textbf{Introduction---} Altermagnetism is a recently identified magnetic symmetry class that hosts ferromagnetic-like macroscopic phenomena within a compensated antiferromagnet \cite{smejkal_crystal_2020, hayami_momentum-dependent_2019, smejkal_anomalous_2022,smejkal_emerging_2022,Smejkal2022-nz,bai_altermagnetism_2024,mcclarty_landau_2024,song_altermagnets_2025,jungwirth_altermagnetism_2025,fender_altermagnetism_2025,jungwirth_symmetry_2026}. MnTe has emerged as a protypical altermagnet \cite{lovesey_templates_2023,osumi_observation_2024,krempasky_altermagnetic_2024, amin_nanoscale_2024,lee_broken_2024,negi_mnte_2025}, notable for exhibiting the anomalous Hall effect (AHE) and spin-split bandstructure despite vanishing net magnetization \cite{wasscher_evidence_1965,gonzalez_betancourt_spontaneous_2023,mazin_origin_2024, kluczyk_coexistence_2024,bey_conductivity_2026,liu_strain-tunable_2025}. A signature of altermagnets such as MnTe is that these signatures of time-reversal breaking are sensitive to the orientation of the Néel vector, \textbf{L}. Consequently, developing efficient methods to tune the AHE by manipulating \textbf{L} is an objective of ongoing research \cite{liu_strain-tunable_2025,smolenski_strain-tunability_2025, liu_observation_2026,liebman-pelaez_strain_2026}. 

\begin{figure} 
   \centering
    \includegraphics[width=\textwidth]{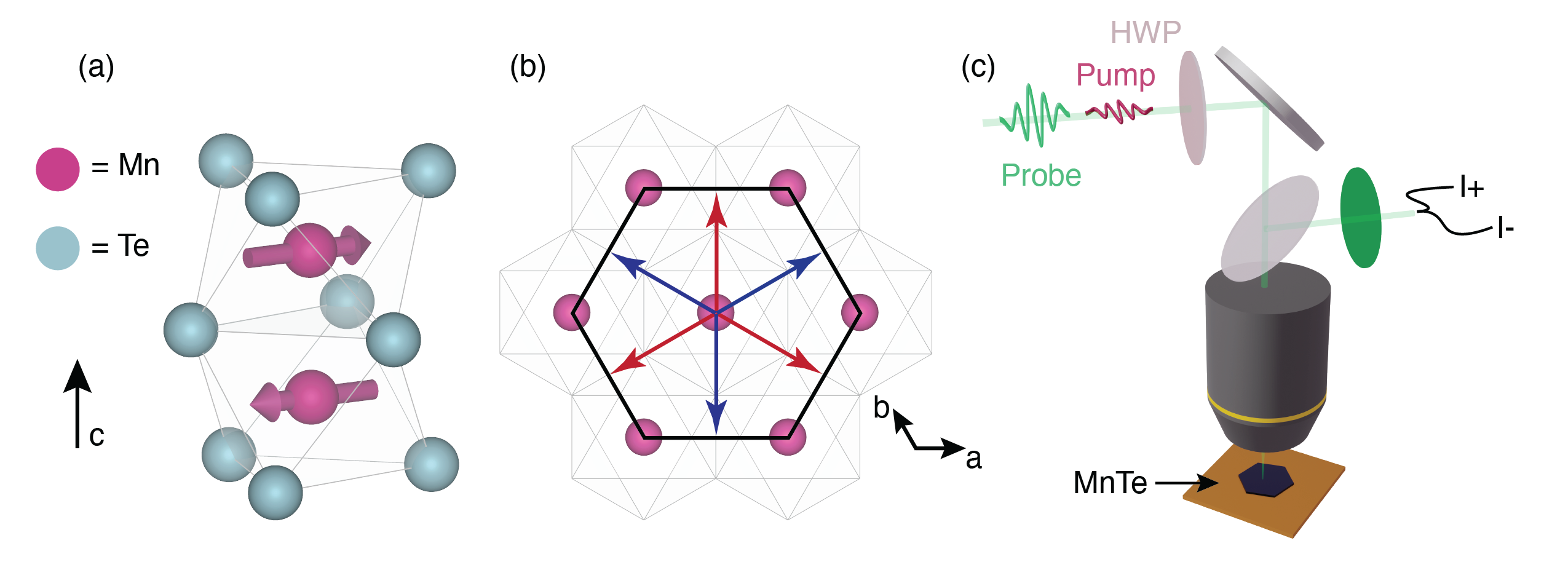}
\caption{Crystal structure side (\textbf{a}) and top (\textbf{b}) view demonstrating hexagonal lattice structure and screw rotation connecting $ \text{Mn}^{2+}$ ions. (\textbf{c}) Schematic of the collinear pump-probe configuration. Pump and probe beams are focused onto the sample via an objective, with the transient polarization rotation measured using a balanced detection scheme.}
    \label{fig:fig1}
\end{figure}

MnTe crystallizes into the hexagonal NiAs-type structure with space group $ P6_3/mmc$ (Fig. \ref{fig:fig1}a) and becomes antiferromagnetic below a Néel temperature ($T_{\rm{N}}$) of $\sim$310 K \cite{Szuszkiewicz2005-wh}. The unit cell contains two $ S = 5/2 $ $ \text{Mn}^{2+} $ sites that are related by 6-fold screw rotation in real space and 180$^\circ$ rotation in spin space \cite{kunitomi_neutron_1964}. The localized moments form A-type antiferromagnetic (AFM) order characterized by ferromagnetic alignment within the basal planes that alternates antiferromagnetically along the c-axis.

The primary order parameter is \textbf{L} \cite{mcclarty_landau_2024}, which here manifests as the difference in the magnetization between adjacent planes. The direction of \textbf{L} within the plane is determined by the competition between two terms in the free energy \cite{liebman-pelaez_strain_2026}, 
\begin{equation}
    F(\varphi) = -K_6 \cos(6\varphi) - K_2 \cos\left[2(\varphi - \varphi_{\varepsilon})\right].
\label{eq:1_free_energy}
\end{equation}
The intrinsic magneto-crystalline anisotropy (MCA, parametrized by $K_6$) respects the 6-fold rotational symmetry, producing six easy directions for \textbf{L}. Competing with $ K_6 $ is the two-fold symmetric magnetoelastic energy, parameterized by $K_2$, which couples \textbf{L} to uniaxial strain oriented along $\varphi_{\varepsilon}$. The relative strength of $ K_2 $ and $ K_6 $ dictates the equilibrium texture of \textbf{L} and, consequently, the magnitude and sign of the AHE and spin-splitting of bands. In the limit that $ K_6 \gg K_2 $, the direction of $\mathbf{L}$ is pinned to one of the symmetry equivalent orientations shown in Fig.~\ref{fig:fig1}b. Conversely, when $ K_6 \ll K_2 $, these directions are no longer local minima of $ F( \varphi) $, and \textbf{L} is pinned instead by the local strain environment. 
 
Previously, an approximate upper bound on $K_6$ in MnTe was obtained by measurements of antiferromagnetic resonance (AFMR) using THz transmission spectroscopy \cite{povarov_low-energy_2025}. The resonant peak corresponds to the zero-wavevector mode of the lower frequency of the two spin wave branches of MnTe. The AFMR frequency, $\Omega$, was observed to be linear in magnetic field applied along the $c$-axis $(B_c)$ in a frequency range from 30-350 GHz and $B_c$ in the range 1-13 T. Within experimental uncertainty, the extrapolation to $B_c=0$ was consistent with $\Omega=0$, which signals a gapless Goldstone mode arising from continuous rotational symmetry (i.e., $K_6=0$) in the basal plane. However, since the AFMR measurements were performed in a large magnetic field, the zero-field gap frequency remained undetermined. 

In this work, we combine equilibrium and time-resolved polarimetry to characterize the free-energy landscape in bulk crystalline MnTe. The time-resolved measurements extend the determination of the AFMR frequency to zero magnetic field, hence directly measuring the anisotropy-induced gap. In addition to accessing the zero-field limit, the spatial resolution of the optical probe allows us to map variations of the spin-wave gap with location on the crystal, revealing a distribution of frequencies that extends from $\approx$ 10 to 30 GHz. Using simultaneous mapping of the equilibrium orientation of the \Neel vector, we are able to place an upper bound on the intrinsic magnetocrystalline anisotropy and to demonstrate that both the equilibrium and dynamical properties of $\mathbf{L}$ are determined almost exclusively by extrinsic strain. Additionally, the dual mapping enables us to determine the mechanism by which laser pulses generate coherent spin-wave oscillations.  Taken together, our results suggest that the direction of \Lvec, and therefore the spin-splitting of the bands and the AHE, can be effectively tuned by strain applied either in the plane or perpendicular to it.

\textbf{Results---}The onset of \Neel order below $T_{\rm{N}}$ spontaneously lowers the rotational symmetry to $C_2$, inducing principal optic axes aligned parallel and perpendicular to $\mathbf{L}$. As shown in Refs.\cite{Liebman-Pelaez2026-rq,liebman-pelaez_strain_2026}, the local direction of $\mathbf{L}$ can be determined by optical polarimetry.  Here we show that photoexcitation of MnTe generates coherent oscillations of $\mathbf{L}$ that correspond to the $q\approx 0$ mode of the acoustic magnon branch (in which the oscillations are primarily in the basal plane). Time-resolved polarimetry probes these oscillations via the arrangement shown schematically in Fig. \ref{fig:fig1}c. Pump (700 nm) and probe (520 nm) beams are spatially overlapped and focused onto the sample at normal incidence using an objective lens (N.A.= 0.25).  After reflection, the time-delayed probe beam is directed into a balanced photodetection scheme that records the transient polarization rotation.  The details of extracting the orientation of Néel vector as a function of time from the balanced detector signal are presented in Supplemental Information.  
\begin{figure*}[t] 
    \centering
    \includegraphics[width=\textwidth]{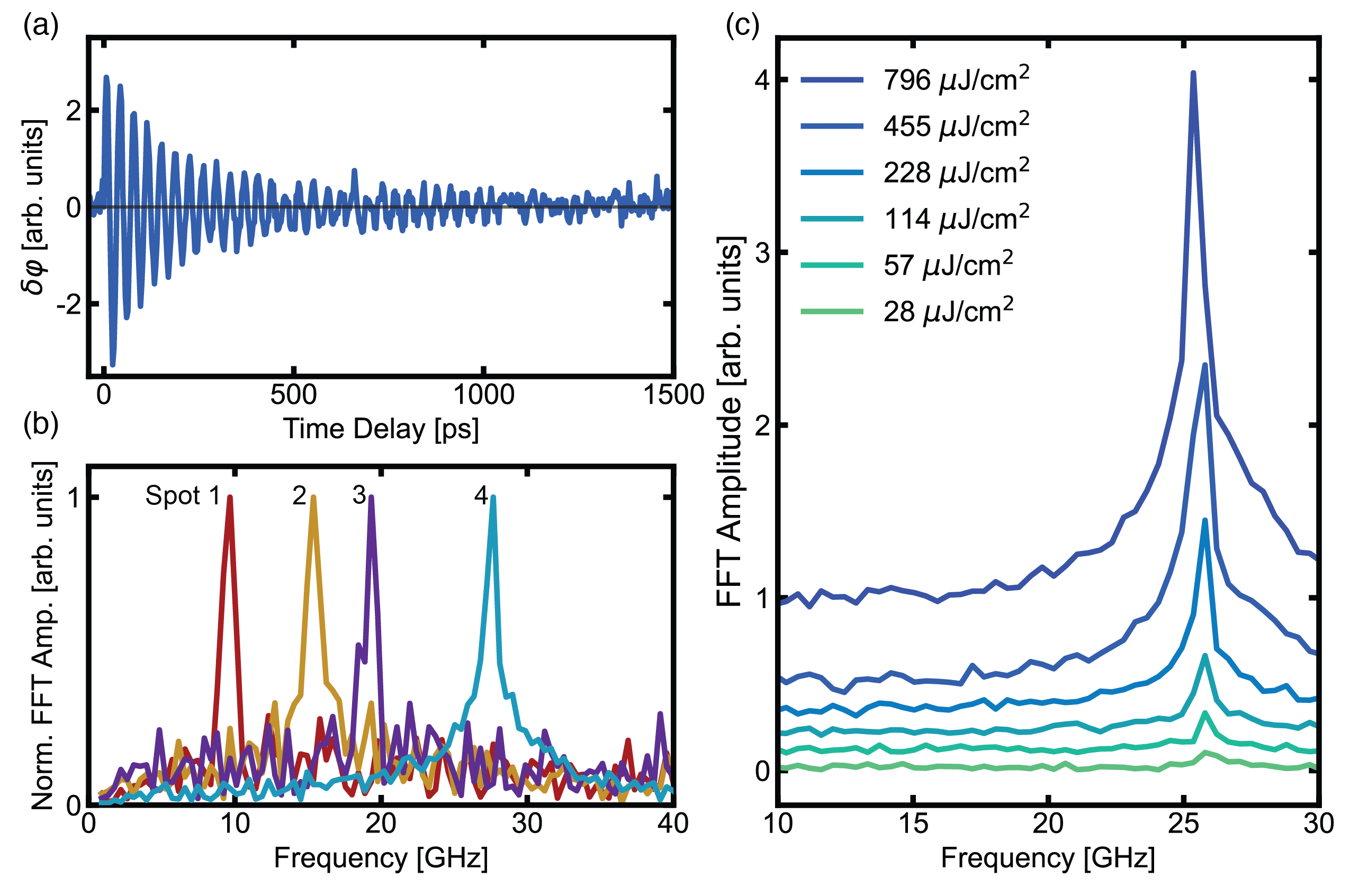}
    \caption{\textbf{Low-temperature magnon dynamics.} (\textbf{a}) Time-resolved pump-probe trace showing the oscillation of \textbf{L} at 10 K. The oscillations are centered on the local equilibrium orientation $\varphi_L(\bf{r})$. (\textbf{b}) Fast Fourier Transform (FFT) amplitude spectra taken at various sample locations revealing a spatial distribution of the magnon gap frequencies. (\textbf{c}) FFT's as a function of pump fluence measured at a single fixed location at 20 K. The frequency remains constant over the full range of applied fluence.}
    \label{fig:fig2}
\end{figure*}

The dynamics of photogenerated spin waves in MnTe show a pronounced contrast between low and high temperature regimes, with a gradual crossover that occurs at $T\approx 120$ K. Figs.~\ref{fig:fig2}a-c highlight the key features of the dynamics at $T=10$ - 20 K that change dramatically when the temperature is raised. First, as shown in Fig.~\ref{fig:fig2}a, the low $T$ oscillations of \Lvec are centered on its local equilibrium orientation, $\varphi_L(\mathbf{r})$. Fig.~\ref{fig:fig2}b shows the resonance sampled at various locations on the crystal, revealing a distribution of frequencies in the range $\approx 10$ - 30 GHz. Through simultaneous mapping of the resonant frequency and the orientation of \Lvec, we find that $\Omega(\mathbf{r})$ and $\varphi_L(\mathbf{r})$ are uncorrelated at low $T$. Finally, as seen in Fig.~\ref{fig:fig2}c, $\Omega$ does not change as the pump fluence, $\Phi$, increases.  

\begin{figure*}[t] 
    \centering
    \includegraphics[width=\textwidth]{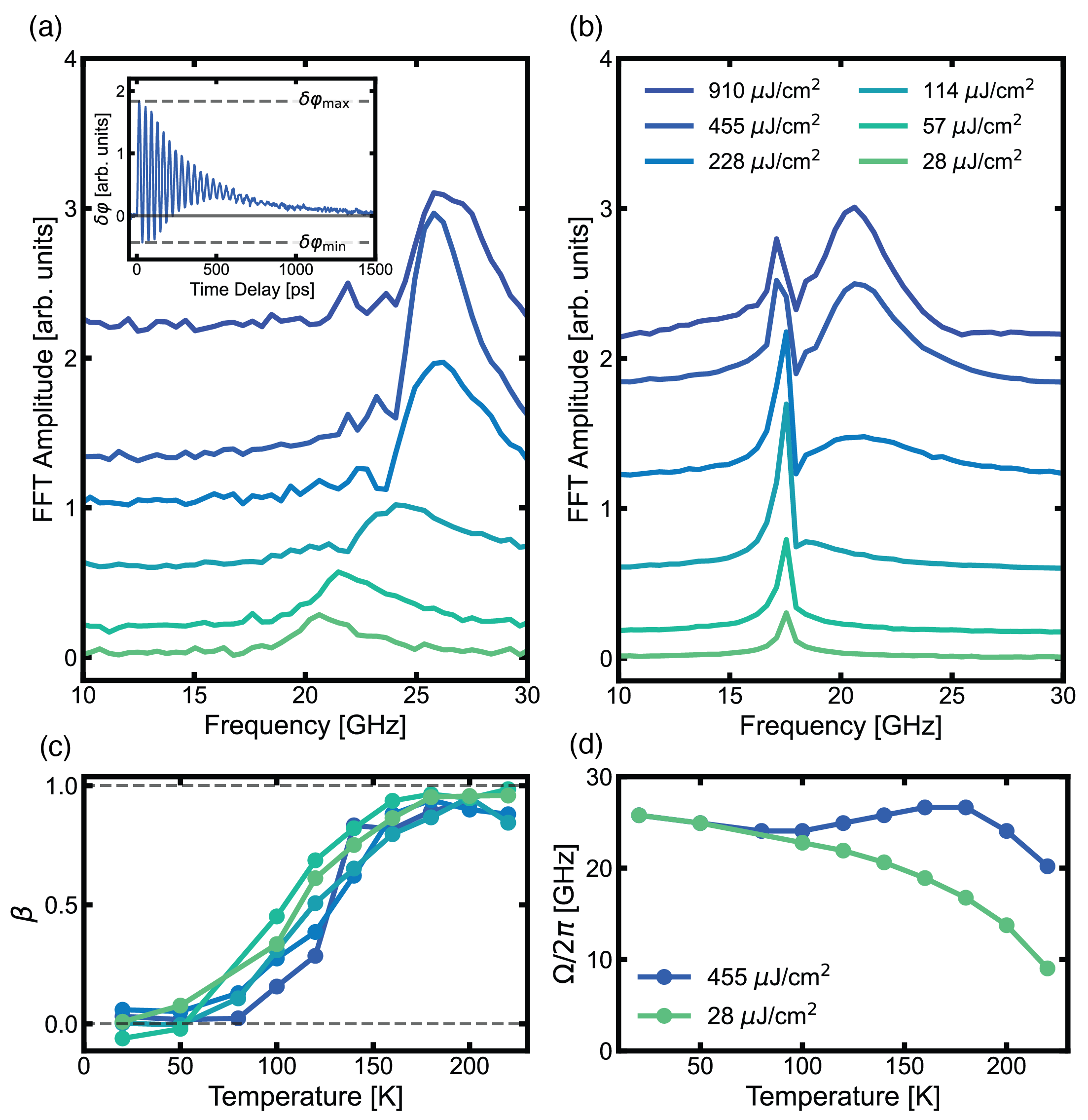}
    \caption{\textbf{High-temperature dynamics: gap frequency is tunable with laser fluence.} FFT amplitude spectra as a function of laser fluence when \textbf{L} is oriented near an (\textbf{a}) easy axis and (\textbf{b}) a hard axis at 140 K. Both locations exhibit a peak whose frequency increases with pump fluence. The inset of (\textbf{a}) reveals qualitatively different time-dependence in this regime, as the center of the \textbf{L} oscillation is displaced from the equilibrium position. (\textbf{c}) Temperature dependence of the displaced center of oscillation, illustrating a crossover from an impulsive ($\beta = 0$) to a step-like ($\beta = 1$) excitation regime as the temperature increases. (\textbf{d}) Temperature dependence of $\Omega$ at a low and high fluence showing the onset of $\Phi$-dependent frequency near 100 K.}
    \label{fig:fig3}
\end{figure*}

Fig.~\ref{fig:fig3} illustrates the contrast in photogenerated spin-wave dynamics when the temperature is raised above $\approx 120$~K. Figures~\ref{fig:fig3}a 
and \ref{fig:fig3}b show the fluence dependence of the resonance measured at locations where $\mathbf{L}$ is close to the easy and hard axes, respectively. At both locations, the spectra show a strong dependence on $\Phi$; in each case, there is a mode whose frequency increases with increasing fluence. This trend 
is unexpected, as typically greater pump fluence increases the transient temperature, reduces spin stiffness and lowers resonance frequencies \cite{mondal_laser_2018,xia_manipulation_2024}. Furthermore, the nature of the oscillations is markedly 
different in the two temperature regimes. As shown in the inset to Fig.~\ref{fig:fig3}a, the center of oscillation is now displaced to a non-equilibrium orientation 
of $\mathbf{L}$. Fig.~\ref{fig:fig3}c illustrates the $T$-dependence of this offset through a plot of the parameter $\beta \equiv (\delta\varphi_\text{max} + 
\delta\varphi_\text{min})/(\delta\varphi_\text{max} - \delta\varphi_\text{min})$, 
which normalizes the shift by the oscillation amplitude.

A further contrast with the low $T$ dynamics is the emergence of a critical dependence of the photogenerated spin waves on the local orientation of \Lvec. Fig.~\ref{fig:fig3}a shows that when the equilibrium direction of \Lvec is close to an easy axis, the shift of $\Omega(\mathbf{r})$ towards higher frequency is a continuous function of $\Phi$. The $\Phi$ dependence measured at locations where $\varphi_L(\mathbf{r})$ is near a hard axis (Fig.~\ref{fig:fig3}b) is clearly very different.  The sharp mode seen in the low $\Phi$ limit persists as $\Phi$ increases, while a second higher frequency resonance appears only above a threshold fluence. 
\begin{figure*}[t]
 \centering
    \includegraphics[width=\textwidth]{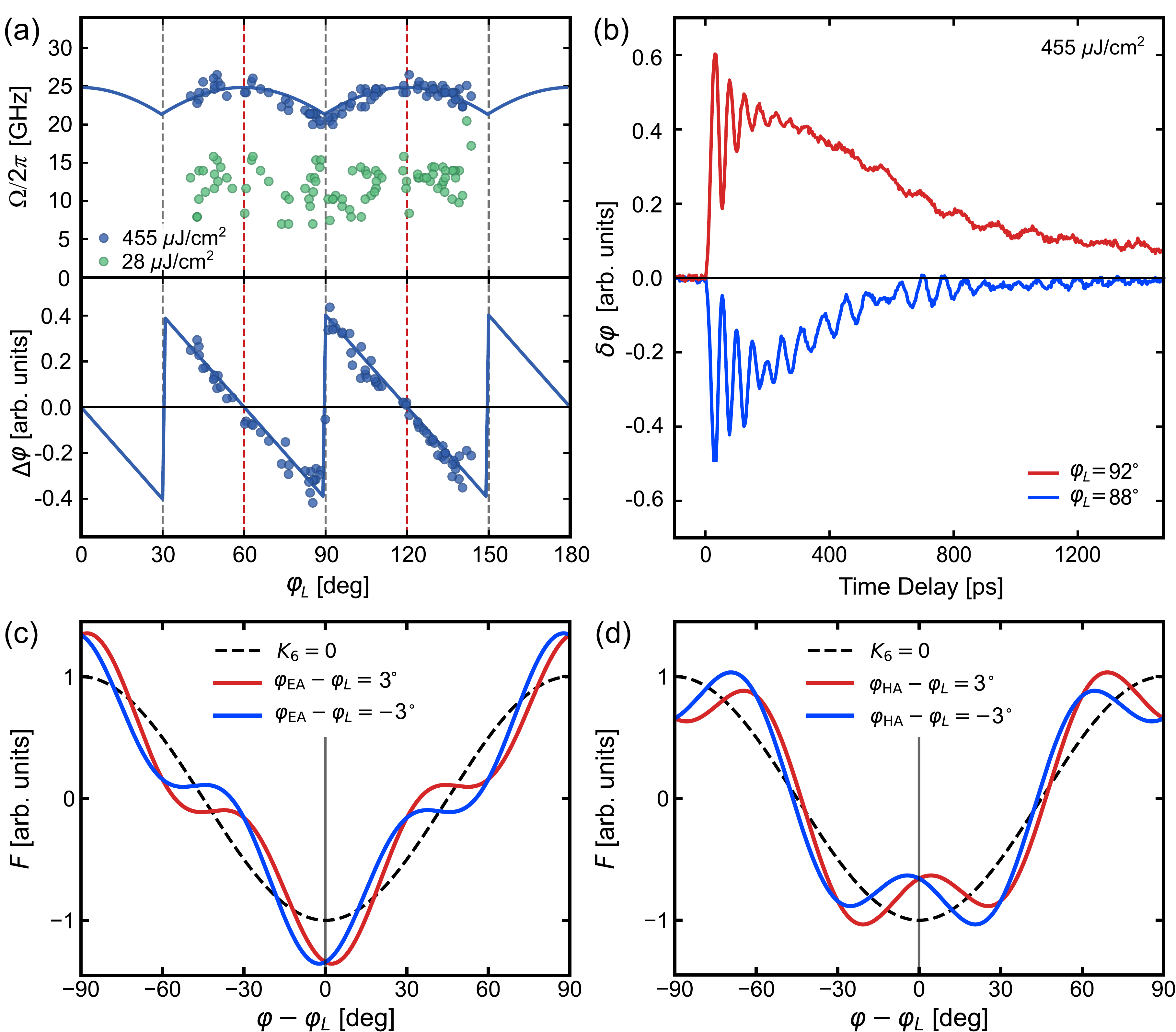}
    \caption{\textbf{Photo-induced enhancement of $K_6$.} (\textbf{a}) Gap frequency, $\Omega(\bf{r}$, (top) and pump-induced displacement (bottom), $\Delta\varphi(\bf{r})$, plotted against $\varphi_L(\bf{r})$ at 180 K for pump fluences 28 $\mu$J/cm$^2$ and 455 $\mu$J/cm$^2$. Red and gray dashed lines represent easy (EA) and hard (HA) axes, respectively. Under the high pump fluence, the magnon frequency exhibits a cycloid-like variation with $\varphi_L$, with maxima at the EA. $\Delta\varphi(\varphi_L)$ also shows a 6-fold symmetric pattern, in this case a sharp sawtooth. The sign of $\Delta\varphi$ indicates that the pump drives $\bf{L}$ towards the EA. (\textbf{b}) Magnon oscillations measured at 180 K and 455 $\mu$J/cm$^2$ at locations where $\varphi_L$ is on either side of, but within $\sim2^{\circ}$ of the HA. The pump-induced displacement abruptly changes sign across the HA. (\textbf{c,d}) Energy landscape change induced by $K_6$ enhancement. The pre-excitation landscape is given by the black dashed curve with $K_6=0$. The post-excitation landscape is illustrated assuming a ratio $K_6/K_2=0.36$, with blue and red curves representing $\varphi_L=\pm3^{\circ}$ with respect to the EA in (\textbf{c}) and the HA in (\textbf{d}).
    }
    \label{fig:fig4}
\end{figure*} 

The dependence of the dynamics on the direction of $\mathbf{L}$ is clarified in the top panel of Fig.~\ref{fig:fig4}a, where each point represents $\Omega(\mathbf{r})$ plotted against $\varphi_L(\mathbf{r})$. The two sets of data points correspond to $\Omega$ measured at low and high values of pump laser fluence. In contrast to the low-$T$ regime, a clear correlation between $\Omega(\mathbf{r})$ and the 
local direction of $\mathbf{L}$ emerges above a threshold fluence; here, $\Omega(\varphi_L)$ exhibits a cycloid-like pattern with cusp-like minima when $\mathbf{L}$ is parallel to a hard axis. Moreover, as shown in the lower panel of Fig.~\ref{fig:fig4}a, there is a striking correlation between the initial photoinduced displacement of the Néel vector, $\Delta \varphi$, and $\varphi_L$. 
The $(\Delta \varphi, \varphi_L)$ pairs fall accurately on a sharp sawtooth pattern, featuring continuous zero-crossings when $\mathbf{L}$ aligns with an easy-axis (next-nearest-neighbor $\text{Mn}^{2+}$, see Fig.~\ref{fig:fig1}b) and singular zero-crossings near hard-axis orientations (nearest-neighbor $\text{Mn}^{2+}$). Fig.~\ref{fig:fig4}b highlights the discontinuity in the photogenerated spin waves on crossing a hard-axis orientation by showing the time-domain response at locations where $\varphi_{L}$ is within $2^\circ$ of the nearest-neighbor direction, but on opposite sides, revealing an abrupt phase shift of $\pi$.

\textbf{Discussion---}

\textit{Low T regime}: The observation that at low temperature $\mathbf{L}$ oscillates about its equilibrium orientation suggests that the perturbation that initiates the spin precession is impulsive. The frequency of the ensuing oscillations is determined by the local curvature of the equilibrium free energy, that is,

\begin{equation}
\Omega^2(\varphi_L,\varphi_\varepsilon)\propto\frac{\partial^2F}{\partial\varphi^2}|_{\varphi_L}  \propto K_2 \cos[2(\varphi_L - \varphi_\varepsilon)] + 9K_6 \cos(6\varphi_L),
\label{eq:2_curvature}
\end{equation}
where $\varphi_\varepsilon$ is the direction of the local strain. Despite the nominal dependence of $\Omega$ on $\varphi_L$ expressed in Eq.~\ref{eq:2_curvature}, our measurements show that at low $T$ the local spin-wave frequency and direction of \Lvec are not correlated.  We therefore conclude that the distribution of spin-wave frequencies shown in Fig.~\ref{fig:fig2}b is consistent with $K_6=0$, in which case $\varphi_L=\varphi_\varepsilon$ and the spread in $\Omega(\mathbf{r})$ reflects the inhomogeneity of the local strain parameter, $K_2$. 

Given the presence of strain inhomogeneity we cannot determine $K_6$ with precision. However, we can obtain an upper bound by reasoning as follows.  At locations on the crystal where $\varphi_L$ is near an easy axis it is possible that $K_2$ is zero and the magnitude of $\Omega$ reflects entirely the contribution from $K_6$. From measurements at such locations at 10 K, we obtain an upper bound on the intrinsic spin-wave gap, $\Omega_b$, of $2\pi\times 15$ GHz. 
The magnitude of the gap, together with the spin wave velocity, $v_s$ determine the intrinsic correlation length for spin-orientation as well as the characteristic domain wall width, through the relation \cite{hubert_magnetic_1998},
\begin{equation}\xi = \frac{v_{s}}{\Omega}.
\end{equation}
Combining our determination of $\Omega_b$  with $v_s\approx 17000$ m/s as measured by inelastic neutron scattering (\cite{Liu2024}), we obtain $\xi \gtrsim 180$ nm.  This lower bound on the wall width is on the order of several hundred lattice constants, implying that domain walls are not susceptible to local pinning on the atomic scale. In a strain-free regime domain walls, although very wide, would be aligned with the crystalline axis, but would be expected to be highly mobile in the presence of externally imposed free energy gradients.  
\\

\noindent\textit{High T regime}: The displacement of the center of oscillations at high $T$ implies a crossover from an impulsive to a step-like perturbation that persists throughout the magnon lifetime. Because this perturbation is long-lived, the subsequent dynamics are shaped by a non-equilibrium free-energy landscape. Below, we demonstrate that the essential features observed at high $T$ and $\Phi$---namely, the sawtooth pattern of $\Delta \varphi(\varphi_L)$ and the cycloid-like structure of $\Omega(\varphi_L)$---are direct consequences of this non-equilibrium $F(\mathbf{r})$.

First, recall that the two zero crossings of the sawtooth shown in Fig.~\ref{fig:fig4}a occur when $\mathbf{L}(\mathbf{r})$ is oriented along one of the six symmetry axes. This observation implies that spin waves are generated by photomodulation of $K_6$; uniquely at these orientations, a photoinduced $\delta K_6$ leaves the equilibrium direction of $\mathbf{L}$ unchanged. Conversely, at locations where strain aligns $\mathbf{L}$ away from a symmetry axis, $\delta K_6$ shifts the free-energy minimum, which triggers the resulting precessions. This framework also accounts for both the singular change in amplitude when the equilibrium $\mathbf{L}$ is near a hard axis, and the continuous change when $\mathbf{L}$ is close to the easy axis. 

The distinction between the two zero crossings is illustrated in Fig.~\ref{fig:fig4}c and \ref{fig:fig4}d, which depict $F(\varphi)$ for equilibrium orientations of \Lvec close to an easy and hard axis respectively. Fig.~\ref{fig:fig4}c and \ref{fig:fig4}d each compare the free energy assuming that $K_6=0$ (dashed line) with $F(\varphi)$ after photoexcitation induces $\delta K_6=0.36K_2$. The red and blue curves correspond to equilibrium orientations that differ from the high symmetry axes by $\pm 3^\circ$.  The contrast in the nonequilibrium free energy landscape between the hard and easy axes is evident.  In particular, it is clear that $\delta K_6$ induces a singular change in sign and amplitude of spin-precession when $\varphi_L$ is proximate to the hard axis, and a continuous sign change for equilibrium orientations near the easy axis. 

This qualitative picture of the high-$T$ regime can be made quantitative by including a step-like photoinduced enhancement $\delta K_6(t)$ in the free energy shown in Eq.~\ref{eq:1_free_energy}. We obtain $\Delta\varphi(\varphi_L)$ from the shift in the free energy minimum induced by $\delta K_6$ and $\Omega(\varphi)$ from the curvature evaluated at this minimum. The solid lines in Figs.~\ref{fig:fig4}a are fits to $ \Delta \varphi(\varphi_L)$ and $\Omega(\varphi_L)$, respectively, with parameters $K_6=0 $ and $\delta K_6 /K_2= 0.36$.  The model clearly captures the sawtooth pattern of $ \Delta \varphi(\varphi_L)$ and the cycloid-like structure of $\Omega( \varphi_L)$. The persistent low frequency peak seen in Fig.~\ref{fig:fig3}b arises from low $\Phi$ excitation in the tails of the Gaussian laser focal spot. Although we find an excellent fit assuming $K_6=0$, as discussed previously we determine only an upper bound on this parameter. However, it is important to note that our scans readily find locations on the crystal where \textbf{L} is pinned along the magneto-crystalline hard axis. This observation alone establishes a strict upper bound on $K_6/K_2$ at these locations, as there is no stable free energy minimum in the hard axis direction when $ K_6$ is larger than $K_2/9$. This upper bound, $ K_6 \leq K_2/9 $ supports a picture in which local strain-induced anisotropy generically dominates over MCA in MnTe crystals.

\textbf{Conclusions---} We reported spatially resolved measurements of the zero-wavevector spin-wave gap in crystals of the altermagnet MnTe using time-resolved polarimetry.  In the limit of low temperature, we observed that the gap frequency, $\Omega$, varies with location on the crystal in a range from about 10 to 30 GHz. In this regime, $\Omega$ is uncorrelated with the local direction of the \Neel vector and we concluded that its variation is consistent with zero magnetocrystalline anisotropy ($K_6$) and fluctuations in "built-in" local strain. In addition, by considering the value of $\Omega$ at locations where $\mathbf{L}$ points along the easy axis, we place an upper bound of 15 GHz on the intrinsic spin-wave gap.  Finally, we noted that the \textit{upper} bound on the intrinsic value of $\Omega$ leads to a \textit{lower} bound on the domain wall width of approximately 400 lattice constants.

In a regime of higher $T$ and $\Phi$ the amplitude of photoexcited spin waves becomes strongly correlated with the local equilibrium orientation of \Lvec.  The correlation appears when the impulsive perturbation induced by the laser pulse crosses over to a step-like persistent perturbation above 120 K.  In this regime the amplitude of photogenerated spin waves crosses zero when the equilibrium \Lvec orients along a symmetry axis, from which we conclude that the photogeneration mechanism is a transient enhancement of $K_6$.  Although the origin of the photoinduced enhancement of $K_6$ is not known, we speculate that the exponential profile of laser absorption induces strain in the direction normal to the surface, enhancing the MCA through the symmetry-allowed magnetoelastic coupling, $K_6\propto \epsilon_{zz}L^6$. 
The most significant finding of relevance to potential applications of altermagnets such as MnTe is that the direction of the \Neel vector in equilibrium is not fixed by intrinsic magnetocrystalline anisotropy. Rather the direction of \Lvec is highly sensitive to external strain applied either in-plane or normal to the $c$-axis. Thus strain provides a route to control of the non-relativistic spin splitting and magnitude and sign of the anomalous Hall effect. 

\begin{acknowledgments}
We thank Alex Liebman-Pel\'{a}ez, Yue Sun, and Veronika Sunko for helpful discussions. This research was primarily funded by the Quantum Materials (KC2202) program under the U.S. Department of Energy, Office of Science, Office of Basic Energy Sciences, Materials Sciences and Engineering Division under Contract No. DE-AC02-05CH11231, which supported the experimental and theoretical work at the LBNL and UC Berkeley. J.O. received support from the Gordon and Betty Moore Foundation’s EPiQS Initiative through Grant GBMF4537 to J.O. at UC Berkeley. J.K. received support from the National Science Foundation Graduate Research Fellowship Program under Grant No. 2146752. Any opinions, findings, and conclusions or recommendations expressed in this material are those of the author(s) and do not necessarily reflect the views of the National Science Foundation. N.J.G. and R. B. R. were supported by Army Research Office under Cooperative Agreement Number W911NF-22-2-0173. 
\end{acknowledgments}

\bibliographystyle{apsrev4-2} 
\bibliography{references}

\newpage
\setcounter{figure}{0}
\setcounter{table}{0}

\renewcommand{\thefigure}{S\arabic{figure}}
\renewcommand{\thetable}{S\arabic{table}}
\setcounter{secnumdepth}{2}
\setcounter{section}{0}
\renewcommand{\thesection}{S\arabic{section}}

\section*{Nomenclature}

\noindent $\varphi$: azimuthal angle of $\mathbf{L}$ \\
$\langle \varphi(t)\rangle$: instantaneous mean azimuthal angle of $\mathbf{L}$, defined as the average of the upper and lower envelopes of the oscillation trajectory\\
$\varphi_L$: equilibrium azimuthal angle of $\mathbf{L}$ in the absence of pump excitation\\
$\delta \varphi(t)=\varphi(t)-\varphi_L$: transient deviation of azimuthal angle from $\varphi_L$ \\
$\delta\langle\varphi(t)\rangle=\langle\varphi(t)\rangle-\varphi_L$: pump-induced shift of mean azimuthal angle \\
$\Delta\varphi\equiv\max|\delta\langle\varphi(t)\rangle|$: peak pump-induced shift of mean azimuthal angle \\
$\delta K_6$: photo-induced change in $K_6$

\section{Material Synthesis}

Single crystals of MnTe were grown by the Tin-flux method. Mn pieces (Thermo scientific; 99.9\%), Te shots (Thermo scientific; 99.999\%), and Sn shots (Thermo scientific; 99.9999\%) were loaded in a 5-ml aluminum oxide crucible in a molar ratio of 1:1:20. The crucible was sealed in a fused silica ampoule under vacuum and heated to 960°C over 10 h, homogenized at 960 °C for 12 h, and then cooled to 840°C over 100 h. After reaching 840°C, the excess flux was decanted from the crystals using a centrifuge, leaving behind well-faceted shiny multiple hexagonal single crystals with a few millimeters in dimensions.

\section{Extracting $\mathbf{L}$ information from optical signals}
\label{sec:Biref}
\setcounter{equation}{0}
\renewcommand{\theequation}{S2-\arabic{equation}}

\subsection*{Optical signal}
We can write the reflection Jones matrix of the sample without pump excitation as:

\begin{equation}
\mathbf{M} = 
r_0\begin{pmatrix}
1 & k_0 \\
-k_0 & 1
\end{pmatrix}
+\mathbf{R}(\varphi_{b0})r_0
\begin{pmatrix}
b_0 & 0 \\
0 & -b_0
\end{pmatrix}
\mathbf{R}(-\varphi_{b0}),
\label{eq:reflection_matrix}
\end{equation}
\noindent where $r_0$ is the averaged amplitude reflection coefficient. $b_0$ and $k_0$ represent the contributions from birefringence and polar Magneto-optical Kerr effect (MOKE). $\varphi_{b0}$ represents the principal axis orientation of the birefringence. The rotation matrix $R(\varphi)$ is defined as:
\begin{equation}
\mathbf{R}(\varphi)=
\begin{pmatrix}
\cos\varphi & -\sin\varphi \\
\sin\varphi & \cos\varphi
\end{pmatrix}.
\label{eq:rotation_matrix}
\end{equation}

The linear polarization orientation of the probe $\varphi_{\rm{p}}$ is controlled by rotating the half-waveplate (HWP1) before the sample. When the half-waveplate (HWP2) in front of the Wollaston prism has its fast axis co-rotating with the probe polarization: $\varphi_{\text{HWP2}}=0.5(\varphi_{\text{p}}+\pi/4)$, the E-field at the two sensors of the balanced detector can be expressed as:

\begin{equation}
\begin{pmatrix}
E_{+} \\
E_{-}
\end{pmatrix} =
\mathbf{R} \left(\frac{\pi}{8} + \frac{\varphi_{\rm{p}}}{2}\right)
\begin{pmatrix}
1 & 0 \\
0 & -1
\end{pmatrix} 
\mathbf{R} \left(-\frac{\pi}{8} - \frac{\varphi_{\rm{p}}}{2}\right)\mathbf{M} E_0
\begin{pmatrix}
\cos\varphi_{\rm{p}} \\
\sin\varphi_{\rm{p}}
\end{pmatrix}.
\label{eq:BD}
\end{equation}

\noindent The signal reading of the balanced detector:
\begin{align}
S(b_0,\varphi_{b0},k_0,\varphi_{\rm{p}}) & = G (E_+E_+^*-E_-E_-^*) \nonumber \\
& \approx 2GE_0^2r_0r_0^* \left[ \operatorname{Re}(b_0)\sin(2\varphi_{\rm{p}}-2\varphi_{b0}) +\operatorname{Re}(k_0) \right],
\label{eq:BD2}
\end{align}
\noindent where $G$ is the gain factor of the detector.

Pump pulses induce a transient evolution in the optical properties of the sample: $r(t)=r_0+\delta r(t)$, $b(t)=b_0+\delta b(t)$, $k(t)=k_0+\delta k(t)$, and $\varphi_b(t)=\varphi_{b0}+\delta\varphi_b(t)$. Therefore, the pump-probe signal:

\begin{align}
P(t;\varphi_p)&=S(b(t),\varphi_b(t),k(t),\varphi_{\rm{p}})-S(b_0,\varphi_{b0},k_0,\varphi_{\rm{p}}) \nonumber \\
& = 2GE_0^2r_0r_0^* \left[ \sin(2\varphi_{\rm{p}}-2\varphi_{b0})\operatorname{Re}(\delta b)-2\operatorname{Re}( b_0)\cos(2\varphi_{\rm{p}}-2\varphi_{b0})\delta\varphi_b+\operatorname{Re}(\delta k) \right] \nonumber \\
&+2GE_0^2\delta(r_0r_0^*) \left[ \operatorname{Re}(b_0)\sin(2\varphi_{\rm{p}}-2\varphi_{b0}) +\operatorname{Re}(k_0) \right].
\label{eq:tr_signal}
\end{align}

The last term represents the cross-coupling between the transient reflectivity and DC polarization rotation resulting from birefringence and MOKE effects. By fine-tuning HWP2 to balance the detector at individual $\varphi_\text{p}$, the DC polarization rotation signal is eliminated. As a result, the pump-probe signal can be written as:

\begin{align}
P(t;\varphi_p)&\approx 2GE_0^2r_0r_0^* \left[ \sin(2\varphi_{\rm{p}}-2\varphi_{b0})\operatorname{Re}(\delta b)-2\operatorname{Re}( b_0)\cos(2\varphi_{\rm{p}}-2\varphi_{b0})\delta\varphi_b+\operatorname{Re}(\delta k) \right] 
\label{eq:tr_signal2}
\end{align}

Static birefringence from the optical setup is also compensated during balancing the detector with HWP2 and is therefore omitted from the setup Jones matrix.

\begin{figure*} 
    \centering
    \includegraphics[width=\textwidth]{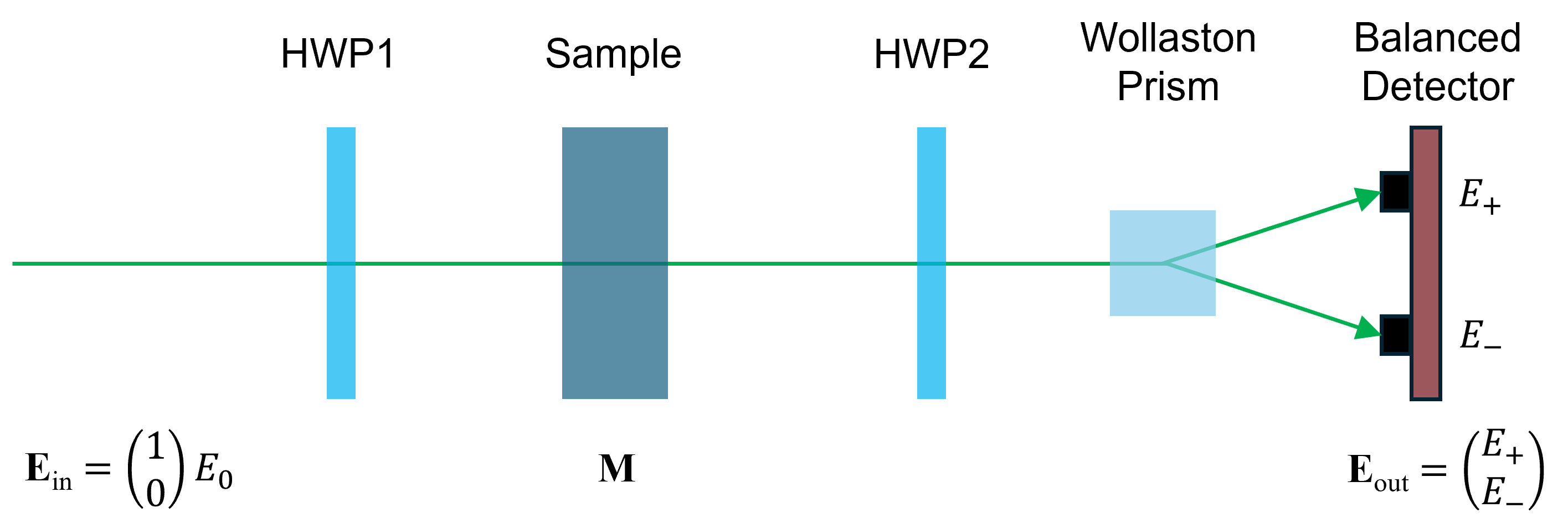}
    \captionsetup{justification=centering}
    \caption{Schematic illustration of the optical setup.}
    \label{fig:figS2-1}
\end{figure*}

\subsection*{Determination of $\varphi_{L}$}

In MnTe, both in-plane $\mathbf{L}$ and in-plane strain break $C_{6z}$ rotation symmetry and therefore contribute to the optical birefringence signal. Recently, Liebman-Peláez, et al. demonstrated that, for a probe wavelength near 520 nm, the amplitude of the optical birefringence remains nearly unchanged under externally applied in-plane strain, while the birefringence principal axis rotates consistently with the expected Néel-vector orientation \cite{liebman-pelaez_strain_2026}. These observations indicate that the strain contribution to the optical birefringence is negligible at this wavelength and that the birefringence is predominantly determined by the Néel order. 

Consequently, the reorientation of birefringence in our time-resolved signal is tied to the reorientation of in-plane $\mathbf{L}$, which occurs at a timescale not faster than the magnon precession ($\sim$50 ps). Therefore, immediately after pump excitation ($t=0^+<1$ ps), $\delta \varphi_b(t)\approx 0$, which leads to:
\begin{align}
P(0^+;\varphi_p)&\approx 2GE_0^2r_0r_0^* \left[ \sin(2\varphi_{\rm{p}}-2\varphi_{b0})\operatorname{Re}(\delta b)+\operatorname{Re}(\delta k) \right].
\label{eq:tr_signal3}
\end{align}

By fitting the measured $\varphi_{\text{p}}$-dependence of $P(0^+)$ to Eq. (\ref{eq:tr_signal3}), we can extract the equilibrium birefringence orientation $\varphi_{b0}$, which serves as an estimation of $\varphi_L$.

\subsection*{Extraction of $\delta \operatorname{Re}(b)$, $\delta\varphi_b$, and $\delta k$}

Once $\varphi_{b0}$ is determined, $\delta\operatorname{Re}(b(t))$, $\delta\varphi_{b}(t)$, and $\delta k(t)$ can be fitted from $\varphi_{\text{p}}$-dependent $P(t)$ based on Eq.(\ref{eq:tr_signal2}). Among them, $\delta\varphi_b(t)$ directly reflects the
time-resolved in-plane rotation of $\mathbf{L}$: $\delta\varphi(t)=\varphi(t)-\varphi_L$, with $\varphi_L$ being the azimuthal angle of $\mathbf{L}$ before pump excitation.

\subsection*{Determination of $\Delta\varphi$ in Fig.\ref{fig:fig4}}

The non-oscillatory component of the $\delta\varphi(t)$ reflects pump-induced displacement in the mean azimuthal orientation of $\mathbf{L}$: $\delta\langle\varphi(t)\rangle$. As exampled in Fig. \ref{fig:figS2-1}, the following procedures are used to estimate $\delta\langle\varphi(t)\rangle$: \\
1) The local maxima and minima of the $\delta\varphi(t)$ trace are identified. \\
2) The upper and lower envelopes, $U(t)$ and $L(t)$, are obtained by piecewise-linear interpolation through the maxima and minima, respectively. \\
3) $\delta\langle\varphi(t)\rangle$ is then estimated by $\delta\langle\varphi(t)\rangle=0.5\left[U(t)+L(t)\right]$.

Finally, the peak pump-induced shift of equilibrium azimuthal angle is estimated by the signed largest excursion of $\delta\langle\varphi(t)\rangle$ in the first 400 ps:

\begin{align}
\Delta\varphi
=
\delta\langle\varphi(t^*)\rangle,
\qquad
t^*=\underset{0<t<400\,\mathrm{ps}}{\arg\max}
\left|\delta\langle\varphi(t)\rangle\right|.
\label{eq:Delta_varphi_eq}
\end{align}


\begin{figure*} 
    \centering
    \includegraphics[width=\textwidth]{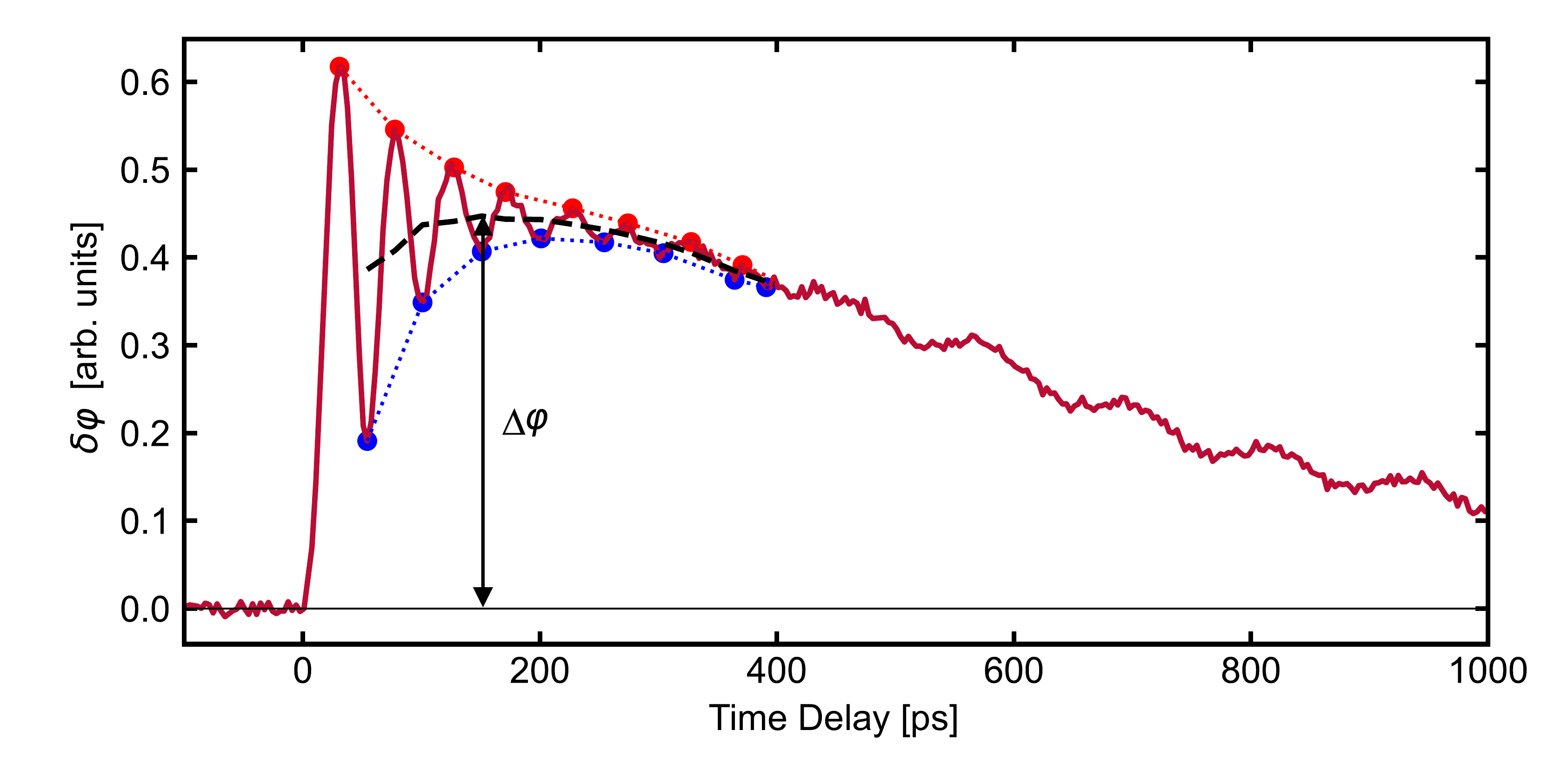}
    \caption{Determination of $\Delta\varphi$ from time traces. Local maxima (red dots) and minima (blue dots) are identified for the first 400 ps. The red (blue) dotted line represents the upper (lower) envelope. The black dashed line is the average of the upper and lower envelopes and serves as an estimation of $\delta\langle\varphi(t)\rangle$, whose peak value gives $\Delta\varphi$.}
    \label{fig:figS2-1}
\end{figure*}

\section{$K_6$-$K_2$ model}
\label{sec:K6K2_model}
\setcounter{equation}{0}
\renewcommand{\theequation}{S3-\arabic{equation}}

As stated in Eq.(\ref{eq:1_free_energy}) of the main text, we consider an in-plane free energy that consists of a 6-fold magneto-crystalline anisotropy contribution and a 2-fold magneto-elastic coupling with the in-plane strain:

\begin{equation}
   F(\varphi)= -K_6\cos(6\varphi)-K_2\cos[2(\varphi-\varphi_{\varepsilon})]
\label{eq:S3-0}
\end{equation}

\subsection*{Equilibrium $\mathbf{L}$ orientation}

The equilibrium $\mathbf{L}$ orientation, denoted as $\varphi_L$, minimizes the free energy:

\begin{equation}
   F(\varphi_L)= \min_{\varphi} F\left(\varphi;\frac{K_6}{K_2},\varphi_{\varepsilon}\right).
\label{eq:S3-1}
\end{equation}

Consequently, $\varphi_L$ can be expressed as a function of the $K_6/K_2$ ratio and $\varphi_{\varepsilon}$: $\varphi_L=g(K_6/K_2,\varphi_{\varepsilon})$. By inverting this relationship, we can alternatively express the strain orientation as: $\varphi_{\varepsilon}=\varphi_{\varepsilon}(K_6/K_2,\varphi_L)$.

\subsection*{Pump-induced $\Delta\varphi$}

Assuming pump pulses induce a step-like change in $K_6$: $K_6^\prime=K_6+\delta K_6$. The new energy landscape has its minimum at $\varphi_L^{\prime}=g(K_6^{\prime}/K_2,\varphi_{\varepsilon})$. The resulting pump-induced displacement in the mean azimuthal angle of $\mathbf{L}$ is:

\begin{align}
   \Delta\varphi&=\varphi_L^{\prime}-\varphi_L = g\left(\frac{K_6^{\prime}}{K_2},\varphi_{\varepsilon}\right)-g\left(\frac{K_6}{K_2},\varphi_{\varepsilon}\right),
\label{eq:S3-2}
\end{align}

By substituting $\varphi_{\varepsilon}=\varphi_{\varepsilon}(K_6/K_2,\varphi_L)$ in to Eq.(\ref{eq:S3-2}), $\Delta\varphi$ can be expressed as:
\begin{align}
\Delta\varphi(\varphi_L)=\Delta\varphi(\varphi_L;K_6/K_2,K_6^{\prime}/K_2). 
\label{eq:S3-3}
\end{align}

The solid curve $\Delta\varphi(\varphi_L)$ in Fig.(\ref{fig:fig4}a) is calculated with $K_6/K_2=0$ and $K_6^{\prime}/K_2=0.36$.

\subsection*{Magnon frequency}

To calculate the magnon frequency, we add the exchange energy and the easy-plane magnetic anisotropy term to the free energy:
\begin{align}
   F_{\text{tot}} =&-\frac{K_6}{2}\left[\cos(6\varphi_{M1})+\cos(6\varphi_{M2})\right]-\frac{K_2}{2}\left[\cos(2\varphi_{M1}-2\varphi_{\varepsilon})+\cos(2\varphi_{M2}-2\varphi_{\varepsilon})\right] \nonumber\\
   &+ J\frac{\mathbf{M}_1\cdot\mathbf{M}_2}{M_s^2}+K_z\frac{M^2_{1z}+M^2_{2z}}{M_s^2},
\label{eq:S3-4}
\end{align}

\noindent where $\varphi_{M1}$ and $\varphi_{M2}$ are the in-plane azimuthal angle of each sub spin. $J$ is the exchange constant. $M_s$ and $K_z$ are the saturation magnetization and easy-plane (hard axis along $z$) magnetic anisotropy energy for each sub-spin lattice, respectively.

Near the equilibrium Néel-vector orientation $\varphi_L$, the in-plane anisotropy energy can be expanded to second order in the angular deviation $\delta\varphi=\varphi-\varphi_L$. The resulting harmonic energy may be expressed in terms of an effective in-plane anisotropy constant $K_{x^{\prime}}$, where the $x^{\prime}$ axis is chosen along the equilibrium Néel vector. The free energy then becomes:
\begin{align}
   F_{\text{tot}} \approx&-K_{x^{\prime}}\frac{M^2_{1x^{\prime}}+M^2_{2x^{\prime}}}{M_s^2} \nonumber\\
   &+ J\frac{\mathbf{M}_1\cdot\mathbf{M}_2}{M_s^2}+K_z\frac{M^2_{1z}+M^2_{2z}}{M_s^2},
\label{eq:S3-5}
\end{align}
\noindent where the in-plane direction $x^{\prime}$ is defined to align with the equilibrium $\mathbf{L}$, having an azimuthal angle of $\varphi_L$. The effective in-plane anisotropy constant: 

\begin{equation}
    K_{x^{\prime}}\approx \frac{1}{4}\left.\frac{\partial^2F}{\partial\varphi^2} \right|_{\varphi=\varphi_L}=9K_6\cos(6\varphi_L)+K_2\cos\left[2(\varphi_L-\varphi_{\varepsilon})\right].
\label{eq:S3-6}
\end{equation}

For a system with free energy in the form of Eq. (\ref{eq:S3-5}), the $\bf{k}=0$ pseudo-Goldstone-mode magnon frequency at zero field can be expressed as \cite{Sun2024}:
\begin{equation}
    \Omega=\sqrt{\omega_x(\omega_J+\omega_x+\omega_z+2\omega_M)},
\label{eq:S3-7}
\end{equation}

\noindent where $\omega_J=2\gamma J/M_s$, $\omega_x=2\gamma K_{x^{\prime}}/M_s$,$\omega_z=2\gamma K_z/M_s$, $\omega_M=4\pi\gamma M_s$. The gyromagnetic ratio $\gamma=1.76\times10^{11}$ rad/($\text{s}\cdot \text{T}$)

From Ref. \cite{Dzian2025}, $\omega_J=2E_J/\hbar=1\times10^{14}$ rad/s, $\omega_z=2E_{\text{an}}/\hbar=3.04\times10^{11}$ rad/s.
The saturation magnetization $M_s$ can be estimated by:
\begin{equation}
    M_s=\frac{gS}{\frac{\sqrt{3}}{2}a^2c}\mu_{\text{B}},
\label{eq:S3-8}
\end{equation}

\noindent where the Landé g-factor $g\approx2$. The total spin quantum number for an Mn$^{2+}$ ion, $S=5/2$. Lattice constants $a=4.14$ \AA, $c=6.75$ \AA \cite{EfremDSa2005}. $\mu_{\text{B}}$ is the Bohr magneton. The above values yield $M_s=462.8$ kA/m. Therefore, $\omega_M=4\pi \gamma M_s=1.0\times10^{11}$ rad/s.

To obtain the observed magnon frequency ($\sim 1\times10^{11}$ rad/s), Eq.(\ref{eq:S3-7}) suggests $\omega_x<10^8$ rad/s. Hence, $\omega_J\gg \omega_z>\omega_M\gg\omega_x$, enabling Eq.(\ref{eq:S3-7}) to be further simplified into:

\begin{equation}
    \Omega\approx\sqrt{\omega_x\omega_J}.
\label{eq:S3-9}
\end{equation}

Substituting $\omega_x=2\gamma K_{x^{\prime}}/M_s$ and Eq.(\ref{eq:S3-6}) into Eq.(\ref{eq:S3-9}) yields:

\begin{equation}
   f = \frac{\Omega}{2\pi}=\eta \sqrt{36\frac{K_6}{K_2}\cos(6\varphi_L)+4\cos\left[2(\varphi_L-\varphi_{\varepsilon})\right]},
\label{eq:S3-10}
\end{equation}

\noindent with:
\begin{equation}
    \eta=\frac{1}{2\pi}\sqrt{\frac{K_2\omega_J\gamma}{2M_s}}.
\label{eq:S3-11}
\end{equation}

\begin{table}[H]
\centering
\caption{Summary of parameter values.}
\begin{tabular}{cc}
\hline\hline
Parameter & Value \\
\hline
$\omega_J$ & $1.00\times10^{14}\,\mathrm{rad/s}$ \\
$\omega_z$ & $3.04\times10^{11}\,\mathrm{rad/s}$ \\
$M_s$ & $462.8\,\mathrm{kA/m}$ \\
$\omega_M$ & $1.00\times10^{11}\,\mathrm{rad/s}$ \\
\hline\hline
\end{tabular}
\end{table}

\subsection*{Pump-induced changes in magnon frequency}

Before pump excitation:
\begin{equation}
   f_0=\eta \sqrt{36\frac{K_6}{K_2}\cos(6\varphi_L)+4\cos\left[2(\varphi_L-\varphi_{\varepsilon})\right]}=f_0\left(\eta,\frac{K_6}{K_2},\varphi_L\right).
\label{eq:S3-11}
\end{equation}

After pump excitation:
\begin{equation}
   f_1=\eta \sqrt{36\frac{K_6^{\prime}}{K_2}\cos(6\varphi_L^{\prime})+4\cos\left[2(\varphi_L^{\prime}-\varphi_{\varepsilon})\right]}=f_1\left(\eta,\frac{K_6}{K_2},\frac{K_6^{\prime}}{K_2},\varphi_L\right).
\label{eq:S3-12}
\end{equation}
Equation (\ref{eq:S3-12}) shows that the $\varphi_L$ dependence of the pump-modified magnon frequency is determined by three parameters. The following values are used for calculating the solid curve in Fig. \ref{fig:fig4}a: $K_6/K_2=0, K_6^{\prime}/K_2 = 0.36,\eta=6$ GHz.

\section{Estimation of the upper bound of $K_6$}
\label{sec:K6_bound}
\setcounter{equation}{0}
\renewcommand{\theequation}{S4-\arabic{equation}}

When $\varphi_L$ is near 6-fold easy axes, 
\begin{equation}
f_0=\eta \sqrt{36\frac{K_6}{K_2}\cos(6\varphi_L)+4\cos\left[2(\varphi_L-\varphi_{\varepsilon})\right]}>\eta \sqrt{36\frac{K_6}{K_2}}.
\label{eq:S4-1}
\end{equation}

Equation (\ref{eq:S4-1}) can be rearranged to give the upper bound of $K_6$:
\begin{equation}
    K_6<\frac{K_2f_0^2}{36\eta^2}= \frac{M_s(2\pi f_0)^2}{18\omega_J\gamma}
\label{eq:S4-10}
\end{equation}

With $\sim$15 GHz magnon frequency observed on spots with $\varphi_L$ near 6-fold easy axes at 10 K, we estimate $K_6<13$ J/m$^3$.

\end{document}